\documentclass[prl,twocolumn,floatfix]{revtex4}
\usepackage{graphicx}
\usepackage{bm}

\newlength{\onecolfig}
\newlength{\twocolfig}
\newcommand{\ion}[2]{\mbox{$^{#2}$#1$^+$}}
\newcommand{\Ca}[1]{\ion{Ca}{#1}}

\newcommand{\Be}[1]{\ion{Be}{#1}}
\newcommand{\lev}[2]{\mbox{#1$_{\mbox{\tiny$#2$}}$}}
\newcommand{\hfslev}[3]{\mbox{#1$^{\mbox{\tiny$#3$}}_{\mbox{\tiny$#2$}}$}}

\newcommand{\unit}[1]{\,\mbox{#1}}

\newcommand{\Hz}{\unit{Hz}}
\newcommand{\kHz}{\unit{kHz}}
\newcommand{\MHz}{\unit{MHz}}
\newcommand{\GHz}{\unit{GHz}}
\newcommand{\THz}{\unit{THz}}

\newcommand{\mW}{\unit{mW}}

\newcommand{\mrad}{\unit{mrad}}
\newcommand{\m}{\unit{m}}

\newcommand{\um}{\unit{$\mu$m}}
\newcommand{\nm}{\unit{nm}}

\newcommand{\mK}{\unit{mK}}
\newcommand{\uK}{\unit{$\mu$K}}
\newcommand{\s}{\unit{s}}

\newcommand{\persec}{\unit{s$^{-1}$}}
\newcommand{\ms}{\unit{ms}}
\newcommand{\us}{\unit{$\mu$s}}

\newcommand{\degree}{\mbox{$^{\circ}$}}

\newcommand{\mT}{\unit{mT}}
\newcommand{\uT}{\unit{$\mu$T}}
\newcommand{\mbar}{\unit{mbar}}

\newcommand{\citesec}[2]{\cite[\S{}#2]{#1}}   % {} to eliminate gap?

\newcommand{\ish}{\mbox{$\sim$}\,}
\newcommand{\ltish}{\protect\raisebox{-0.4ex}{$\,\stackrel{<}{\scriptstyle\sim}\,$}}
\newcommand{\gtish}{\protect\raisebox{-0.4ex}{$\,\stackrel{>}{\scriptstyle\sim}\,$}}

\newcommand{\bra}[1]{\mbox{$\left< #1 \right|$}}
\newcommand{\ket}[1]{\mbox{$\left| #1 \right>$}}

\newcommand{\wee}[2]{\mbox{$\frac{#1}{#2}$}}
\newcommand{\sub}[1]{\mbox{$_{\mbox{\tiny #1}}$}}

\newcommand{\e}[1]{\ensuremath{\times 10^{#1}}} % scientific notation

\begin{document}
\bibliographystyle{apsrev}

%\title{Laser-driven quantum logic gates with precision beyond the fault-tolerant threshold}
\title{High-fidelity quantum logic gates using trapped-ion hyperfine qubits}

\author{C. J. Ballance, T. P. Harty, N. M. Linke, M. A. Sepiol and D. M. Lucas}
\email{d.lucas@physics.ox.ac.uk}

\affiliation{Department of Physics, University of Oxford, Clarendon Laboratory, Parks Road, Oxford OX1 3PU, U.K.}

\date{21 June 2016, v10.2}

% 0. Abstract

\begin{abstract}
We demonstrate laser-driven two-qubit and single-qubit logic gates with fidelities 99.9(1)\% and 99.9934(3)\% respectively, significantly above the $\approx 99\%$ minimum threshold level required for fault-tolerant quantum computation, using qubits stored in hyperfine ground states of calcium-43 ions held in a room-temperature trap. We study the speed/fidelity trade-off for the two-qubit gate, for gate times between 3.8\us\ and 520\us, and develop a theoretical error model which is consistent with the data and which allows us to identify the principal technical sources of infidelity. 
\end{abstract}

\maketitle

%\section{Main text}

% 1. Introductory paragraph

A powerful quantum computer need not require more than a few thousand logical qubits, but the number of physical qubits required depends strongly on the precision with which they can be manipulated~\cite{Steane2003b}. Fault-tolerant quantum error correction typically requires that the errors associated with all operations (qubit initialization, single- and two-qubit logic gates, and readout) must each be below a threshold level of $\approx 1\%$ in order for a quantum computer to function at all~\cite{Knill2005,Raussendorf2007,Fowler2012}. The ability to entangle qubits ``on demand'' has been demonstrated in several physical systems~\cite{Turchette1998,Steffen2006,Isenhower2010,Veldhorst2015} and error rates slightly below threshold have been achieved using trapped ions~\cite{Benhelm2008} and superconducting circuits~\cite{Barends2014}. However, the precision has so far fallen short of that needed for the construction of a practical quantum computer, because error rates at least an order of magnitude below threshold are required for the number of physical qubits per logical qubit to remain reasonable~\cite{Steane2003b,Knill2005,Fowler2012}. The speed of the operations is also an important parameter: gate speed does not need to be fast in absolute terms (a quantum computer derives its power from the exponential scaling of its workspace with the number of qubits, not from its clock speed), but should be sufficiently fast relative to the qubit coherence time that the memory error is also well below the threshold. In general, there is a trade-off between speed and fidelity, both for specific systems (as studied here) and between different platforms. For example, the strong interactions in the solid state permit sub-microsecond two-qubit gates for superconducting qubits, much faster than is typical for trapped ions, but at present also limit qubit coherence times to \ish 100\us. In both these physical systems, the present limitations to gate speed and fidelity are technical rather than fundamental. 

The first detailed proposals for implementing the theoretical ideas of quantum information processing appeared in the 1990s, and were based on laser-cooled trapped ions~\cite{Cirac1995,Steane1997,Wineland1998a}, and on single-electron quantum dots~\cite{Loss1998}. Individual trapped ions possess extremely stable internal states for the storage of quantum information (such states form the basis of some of the most accurate atomic clocks~\cite{Ludlow2015}) and the ion-ion coupling arising from the mutual Coulomb repulsion provides a natural mechanism for implementing multi-qubit quantum logic. As in other physical systems, the quantum logic operations which entangle distinct qubits are the most technically challenging to implement, because --- however stable the internal qubit states --- the quantum information needs to be transmitted between qubits via an external channel which is generally more susceptible to environmental noise. In the case of trapped ions, this channel is the quantized motion of the ions in the harmonic oscillator potential of the trap and is thus sensitive to the effective motional temperature of the ions and to noise in the electric fields used to confine them. The highest fidelity previously reported~\cite{Benhelm2008} for a two-qubit gate in trapped ions was 99.3(1)\% (a level which has recently been equalled using superconducting qubits~\cite{Barends2014}); this used an optical qubit transition and hence required good frequency stability in the optical domain. Qubits based on hyperfine ground states, in common with superconducting qubits, operate in the more convenient microwave domain; in contrast to manufactured solid state qubits, however, the qubit frequency is defined by universal atomic properties and this may simplify large-scale architectures. 

\begin{figure*}[t]
\centering
\includegraphics[width=0.83\onecolfig]{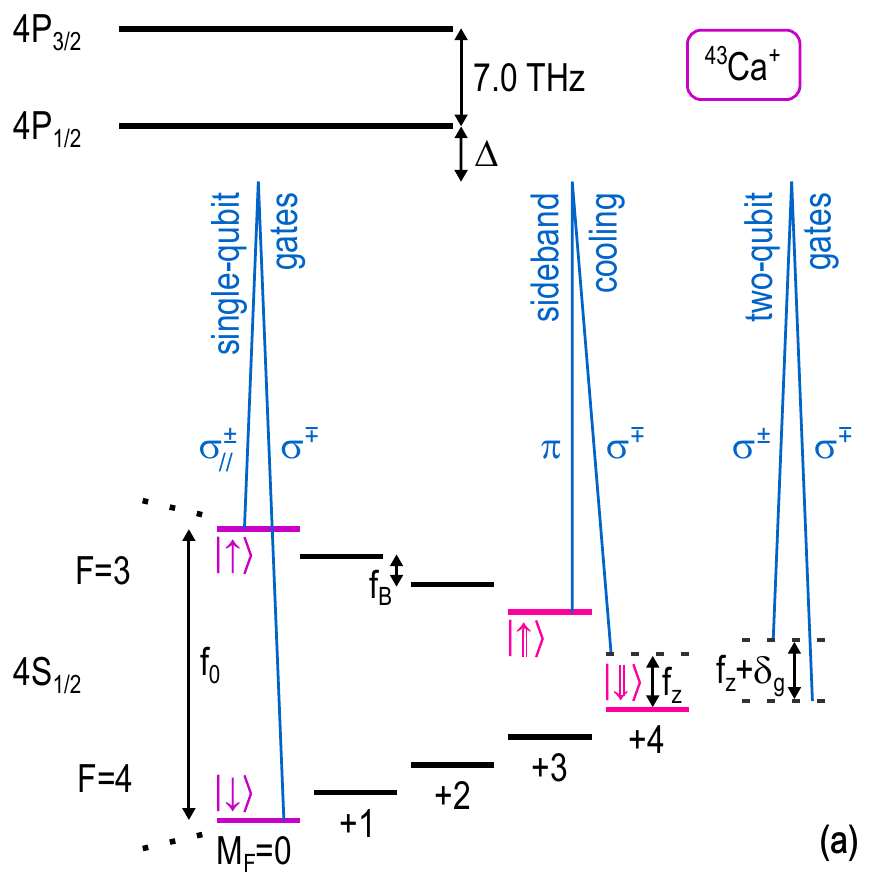} 
%\\
%\vspace{1ex}
\hspace{4em}
\includegraphics[width=0.83\onecolfig]{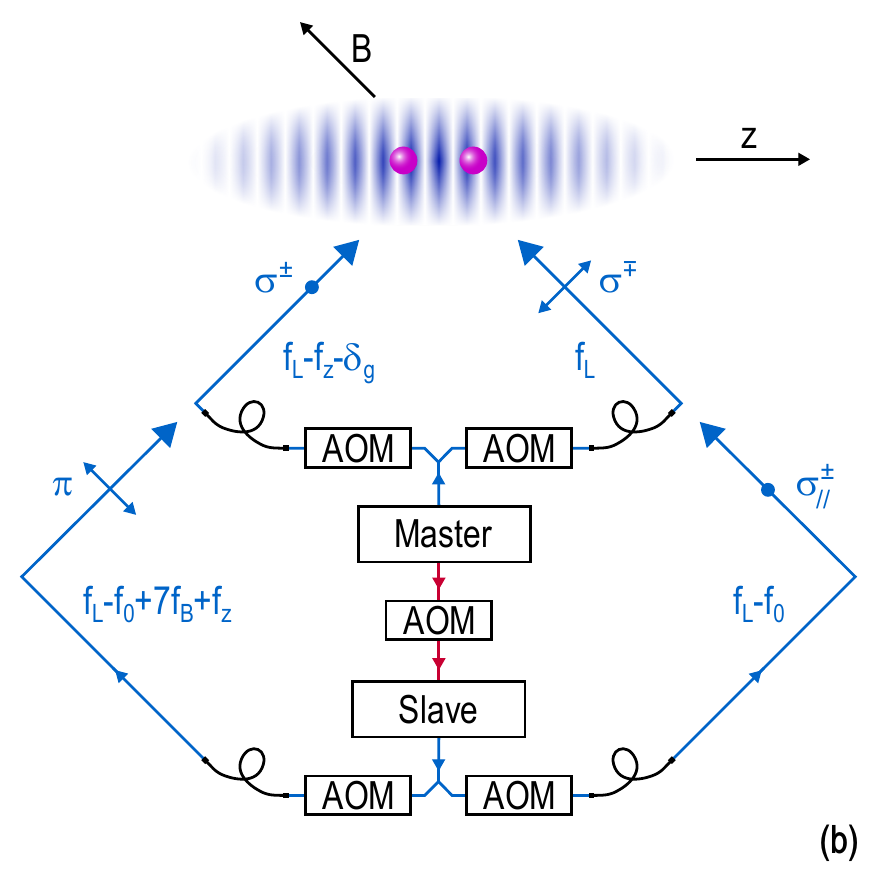}
\caption{
(a) \Ca{43} qubit states and Raman transitions used for sideband cooling, single-qubit and two-qubit gates. The quantization axis is set by a magnetic field $B=0.196\mT$, giving Zeeman splittings $f_B\approx 0.686\MHz$ between adjacent hyperfine states. Raman beams have mean detuning $\Delta\,\ish -1\THz$ from the $\lev{4S}{1/2}\leftrightarrow\lev{4P}{1/2}$ (397\nm) transition. For single-qubit gates, the Raman difference frequency $\delta=f_0$ where $f_0=3.226\GHz$ is the (\ket{\downarrow},\ket{\uparrow}) qubit transition frequency. For sideband cooling, $\delta=f_0-7f_B-f_z$ to cool the axial centre-of-mass motion at $f_z=1.95\MHz$, and $\delta=f_0-7f_B-f_z\sqrt{3}$ to cool the stretch motion. For two-qubit gates, using the (\ket{\Downarrow},\ket{\Uparrow}) states, $\delta=f_z+\delta_g$ where $\delta_g=2/t_g$ with $t_g$ the total gate duration. Global single-qubit $\pi/2$ and $\pi$ rotations used to characterize the two-qubit gate are driven by microwave radiation.
(b) Raman laser system and beam geometry, with the polarizations and frequencies of each beam. The ion separation (3.5\um) is set to be $12\frac{1}{2}$ wavelengths of the travelling standing wave which results from the interference of the two-qubit gate beams. All beams are derived from a master/slave pair of frequency-doubled lasers whose frequency difference ($\approx f_0$) is set by optical injection locking (at 794\nm) via an acousto-optic modulator (AOM)~\cite{Linke2013a}. The beams are frequency-shifted and switched by further AOMs and brought close to the trap using optical fibres. Beam powers are independently stabilized by feedback to the AOMs' r.f. amplitudes. The gate beams are steered onto the ions via mirrors with piezo-electric actuators.
}
\label{F:gatebeams}
\end{figure*}

We report in this Letter an experimental and theoretical study of a two-qubit gate operation~\cite{Leibfried2003} for hyperfine trapped-ion qubits driven by Raman laser beams which, together with microwave-driven single-qubit operations, produces a Bell state (a maximally-entangled state) whose fidelity we measure by quantum state tomography. By independent characterization of the single-qubit errors, we infer the error in the gate operation itself. We develop a theoretical error model for the gate, verify the dominant error contributions in auxiliary experiments, and find good agreement with the experimental results. This both confirms the accuracy of our fidelity result, and shows where future work should be focussed. We also measure the average error in laser-driven single-qubit rotations, using randomized benchmarking. For both the single- and two-qubit gates we systematically explore the trade-off between gate speed and error.

The particular two-qubit gate we apply is a $\sigma_z\otimes\sigma_z$ phase gate~\cite{Leibfried2003} ($\sigma_z$ being the Pauli operator), driven by a pair of Raman laser beams at a mean detuning $\Delta$ from an optical atomic resonance, where the qubits are stored in the \ket{\Downarrow}=\hfslev{4S}{1/2}{4,+4} and \ket{\Uparrow}=\hfslev{4S}{1/2}{3,+3} states of the ground hyperfine manifold of \Ca{43} (where the superscripts denote the quantum numbers $F,M_F$; see figure~\ref{F:gatebeams}a). The Raman beams exert a state-dependent force on the ions, which transiently excites their centre-of-mass axial motion when they are in the \ket{\Downarrow\Uparrow} or \ket{\Uparrow\Downarrow} states, in turn giving an overall phase on the two-qubit wavefunction for these two states. To vary the gate time $t_g$ we adjust $\Delta$ while holding the Raman beam intensity constant; smaller $\Delta$ enables a faster gate, at the cost of increased error due to photon scattering~\cite{Ozeri2007}. The Raman difference frequency is $\delta = f_z + \delta_g$ where $\delta_g = 2/t_g$ and the axial trap frequency is $f_z = 1.95\MHz$. The Raman beams propagate at $45\degree$ to the trap $z$-axis, such that their wave-vector difference lies along $z$ (figure~\ref{F:gatebeams}b). We cool both of the ions' axial modes close to the ground state of motion (mean vibrational quantum number $\bar{n}\approx 0.02$) by Raman sideband cooling; the centre-of-mass mode (with effective temperature $\approx 2\uK$), rather than the stretch mode, is used to implement the gate to avoid coupling to the hotter ($\ish 1\mK$) radial modes of motion~\cite{Roos2008b}. The Raman pulses used to implement the gate are shaped in time, to reduce errors due to off-resonant excitation (see Supplemental Material). 

\begin{figure}[t!]
\centering
\includegraphics[width=0.9\onecolfig]{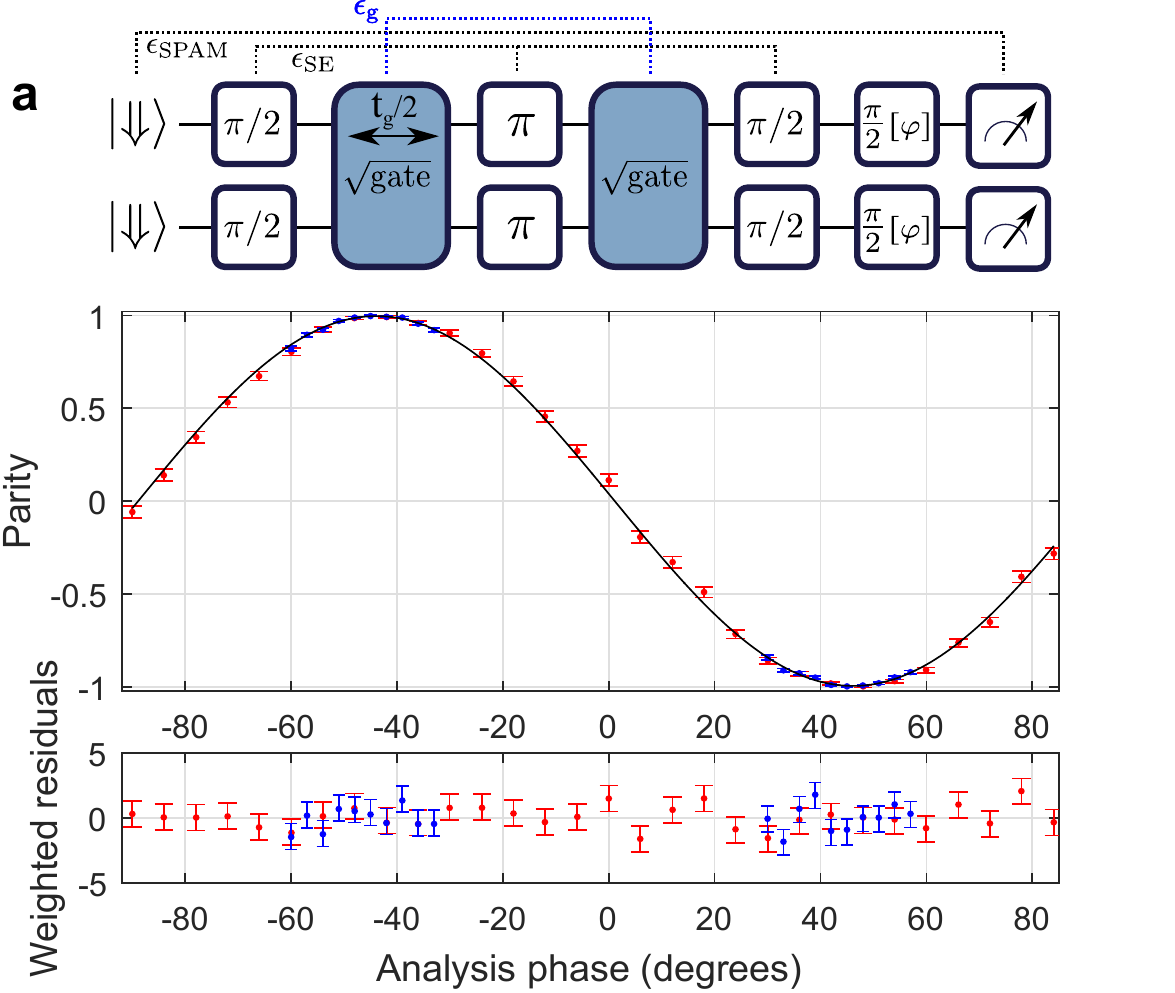} \\
\vspace{3ex}
\includegraphics[width=0.9\onecolfig]{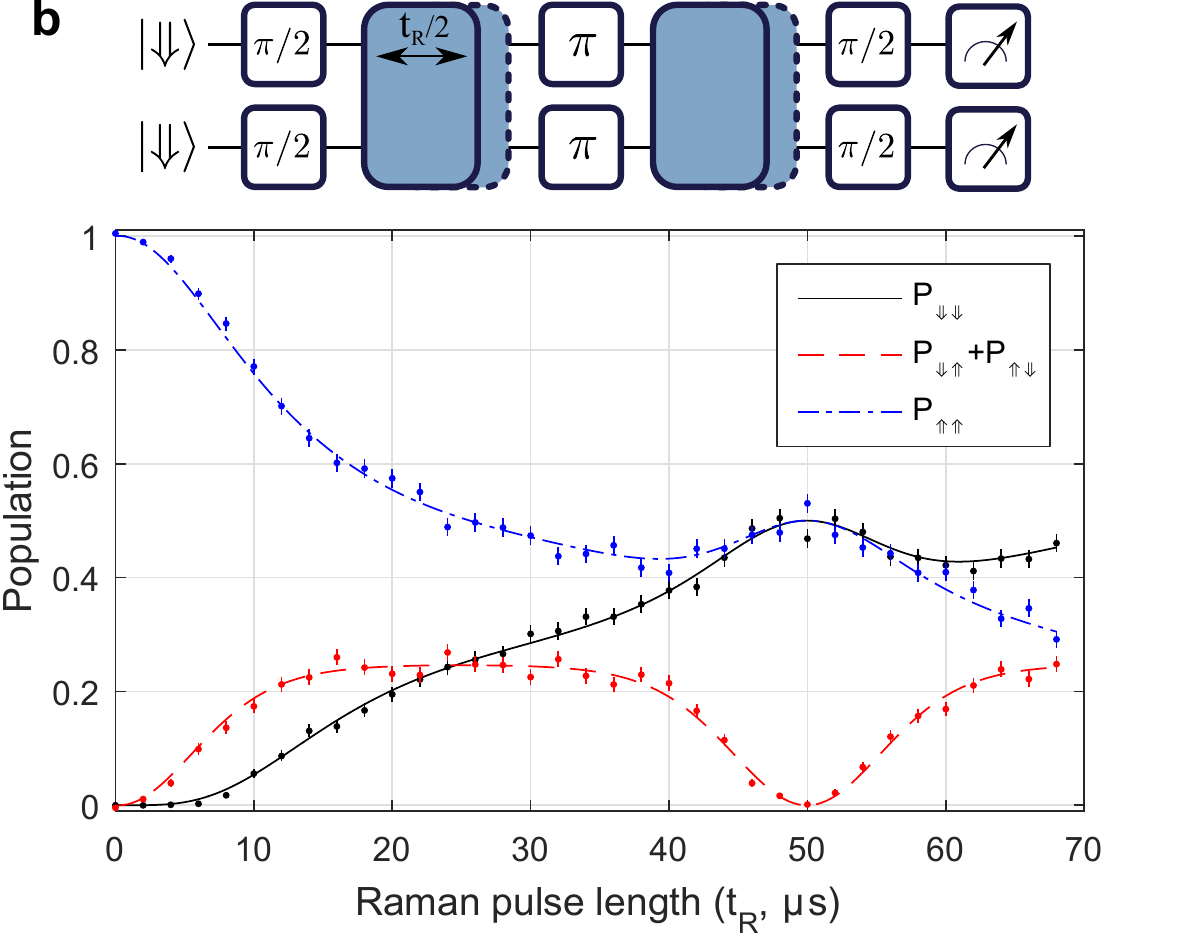}
%\includegraphics[width=0.9\onecolfig]{parityScanWithSequence.pdf}
%\hspace{2em}
%\includegraphics[width=0.9\onecolfig]{dynamicsScanWithSequence3.pdf}
\caption{
(a) Quantum state tomography for the optimum two-qubit gate obtained ($t_g=100\us$). Parity signal ($P_{\Downarrow\Downarrow}+P_{\Uparrow\Uparrow}-P_{\Downarrow\Uparrow}-P_{\Uparrow\Downarrow}$ where $P_\alpha$ is the probability of finding the ions in state $\ket{\alpha}$), obtained by analysing the Bell state with $\frac{\pi}{2}[\phi]$ tomography pulses whose phase $\phi$ is scanned relative to that of the single-qubit spin-echo pulses. The operation sequence is shown above, with the sources of the different contributions $\epsilon$ to the measured Bell state error indicated. Two independent runs are plotted (red: 1000 sequences per point, blue: 2000 sequences per point). The curve is the maximum likelihood fit to the data (see Supplemental Material); the weighted residuals are shown below, and give a reduced $\chi^2=0.87$.  
(b) Population dynamics during the gate operation, obtained by scanning the duration $t_R$ of the Raman laser pulses, as illustrated above. The Bell state is generated at $t_R=t_g=50\us$ here. For pulse durations not equal to integer multiples of $1/\delta_g$ the ions' motion does not return to its initial state, which requires the phase of the force to be synchronized between the two Raman pulses. The curves show the ideal gate dynamics with no free parameters except $t_g$. Error bars in both plots show $1\sigma$ statistical errors, calculated using binomial statistics. 
}
\label{F:parity}
\end{figure}

We divide the gate operation into two pulses, each of duration $1/\delta_g$, embedded within a global spin-echo sequence~\cite{Home2006}, which ideally produces the Bell state $\ket{\psi_+} = (\ket{\Downarrow\Downarrow}+\ket{\Uparrow\Uparrow})/\sqrt{2}$, and use further single-qubit $\pi/2$ rotations to measure the fidelity $F=\bra{\psi_+}\rho\ket{\psi_+}$ of the state $\rho$ obtained~\cite{Leibfried2003}, see figure~\ref{F:parity}a. Thus the measured Bell state infidelity includes both the error $\epsilon_g$ due to the gate operation itself and errors in the single-qubit operations (principally spin-echo $\epsilon\sub{SE}$, state preparation and measurement $\epsilon\sub{SPAM}$). We characterize the single-qubit errors by independent experiments in order to extract the two-qubit gate error; the errors in the single-qubit operations are comparable to or smaller than the gate error over the parameter regime studied (see Supplemental Material). The time-dependent dynamics of the gate operation are in excellent agreement with theory (figure~\ref{F:parity}b).

\begin{figure}[t!]
\centering
\includegraphics[width=\onecolfig]{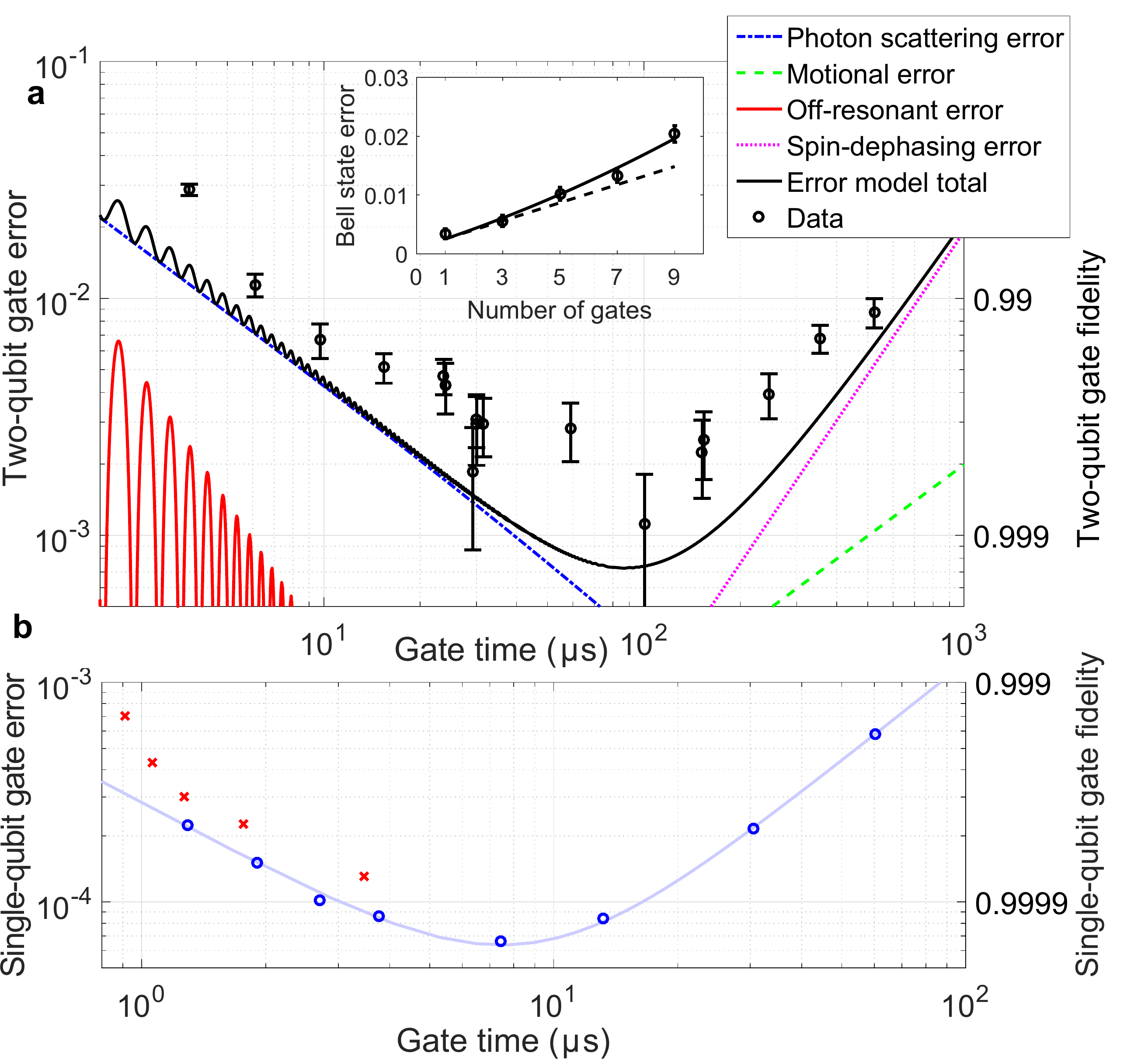}
\caption{%
(a) Measured two-qubit gate error (circles) and error model (lines), plotted against the two-qubit gate duration $t_g$ which was varied by adjusting the Raman detuning $\Delta$ at constant Raman beam power. Single-qubit SPAM and spin-echo errors have been subtracted from the data to allow comparison with the error model (see text). The four largest error contributions are plotted, with their total (black line).
Inset: Bell state error vs number of two-qubit gates, using $t_g=30\us$. The dashed line is the error model prediction of $1.5\e{-3}$ per gate, while the solid line is a quadratic fit allowing for a systematic error in the Raman beam intensity of 0.5\% (consistent with the observed level of drift, see Supplemental Material). 
(b) Measured single-qubit gate error versus gate time $t_{\pi/2}$. The minimum error of 0.066(3)\e{-3} is achieved at $t_{\pi/2}=7.5\us$, at which point the contribution to the error from photon scattering is calculated to be 0.02\e{-3}. Here the gate duration was changed by adjusting the Raman beam intensity; two fixed detunings were used, $\Delta=-1.9\THz$ (circles) and $\Delta=-1.0\THz$ (crosses). Statistical error is smaller than the symbol size. The curve is an empirical model fitted to the data, which allows for differential phase noise between the two Raman beams (see Supplemental Material).
}
\label{F:allgates}
\end{figure}

Results for the complete two-qubit gate data set are shown in figure~\ref{F:allgates}a, where we have normalized throughout for the independently-measured qubit SPAM errors ($\epsilon\sub{SPAM}=1.7\e{-3}$ per qubit) and spin-echo error ($\epsilon\sub{SE}\leq 1.8\e{-3}$). The data are compared with a theoretical model comprising the four leading sources of gate error (see Supplemental Material). The model gives a minimum error estimate since it assumes all control parameters (for example, laser beam intensities) are set to their optimum values; in reality fluctuations in these parameters, and the finite precision with which each can be set, lead to higher errors. Despite this, the data exceed the model prediction by, on average, less than a factor of two over the full range of gate speeds studied. The lowest gate error is found at $t_g=100\us$ (using $\Delta=-3.0\THz$), where the measured Bell state infidelity is $(1-F)=2.5(7)\e{-3}$ after correcting for SPAM error. For this run, the single-qubit spin-echo error contribution is $\epsilon\sub{SE}=1.4(1)\e{-3}$, and we infer a gate error of $\epsilon_g=1.1(7)\e{-3}$, representing more than an order of magnitude improvement compared with that previously reported for hyperfine qubits~\cite{Leibfried2003,Kirchmair2009,Tan2013a}. The measured gate error is consistent with the calculated contributions to $\epsilon_g$ given in table~\ref{T:errorbudget}. The shortest gate time attempted was $t_g=3.8\us$, for which we measure an error $\epsilon_g=29(2)\e{-3}$; this is a five-fold increase in gate speed, and a factor two reduction in error, compared with the fastest previous trapped-ion implementation~\cite{Gaebler2012a}.

We also performed multiple $t_g=30\us$ gates within a single, fixed-length, spin-echo sequence (figure~\ref{F:allgates}a, inset); as $\epsilon\sub{SE}$ and $\epsilon\sub{SPAM}$ do not vary with the number of gates, this allows an independent estimate of the two-qubit gate error. Making the conservative assumption of an error linear in the number of gates, we obtain an error-per-gate of $\epsilon_g=2.0(2)\e{-3}$, independent of all single-qubit and SPAM errors; this value of $\epsilon_g$ is dominated by the calculated increase in photon scattering error compared with the 100\us\ gate. It would be desirable to perform longer sequences of gates, for example for the purpose of randomized benchmarking~\cite{Gaebler2012a}; however, in our system this would not yield useful information on the two-qubit gate error, as the measured error would be dominated by single-qubit errors, due to the effects of 50\Hz\ magnetic field noise and the magnetic field offset between the two ions (see Supplemental Material).

\begin{table}[t!]
\center
\begin{tabular}{|c|c|}\hline
two-qubit gate error source 				& calculated error	\\ 
(at $t_g=100\us, \Delta=-3.0\THz)$		& $/10^{-3}$			\\ \hline
Raman+Rayleigh photon scattering 		& 0.4		   			\\ 
motional heating and dephasing			& 0.2    				\\
spin dephasing 								& 0.2    				\\
Raman beam intensity drift ($<0.5\%$)	& $<0.06$	 			\\
motional temperature ($\bar{n}<0.05$)	& $<0.04$	 			\\
off-resonant effects 						& $<0.01$	 			\\ \hline
total											& 0.9	  				\\ \hline
\end{tabular}
\caption{%
Dominant contributions to the two-qubit gate error $\epsilon_g$, as predicted by our theoretical error model for the conditions under which we measured the lowest gate error $\epsilon_g=1.1(7)\e{-3}$. The uncertainty is at most $\pm1$ in the final digit. For details of the calculations, see Supplemental Material and~\cite{Ballance2014a}. 
}
\label{T:errorbudget}
\end{table}

To test the performance of single-qubit gates ($\pi/2$ rotations) driven by the same Raman laser system, we used instead the qubit states $\ket{\downarrow}=\hfslev{4S}{1/2}{4,0}$ and $\ket{\uparrow}=\hfslev{4S}{1/2}{3,0}$. These states are nearly independent of magnetic field to first order, allowing randomized benchmarking~\cite{Knill2008} to be used (which is necessary to render the gate error observable above the state preparation and measurement errors). The two-qubit gate cannot be directly applied to these ``memory qubit'' states~\cite{Lee2005}, but mapping between similar memory and gate qubits in \Ca{43} has been demonstrated with error 0.2\e{-3} using microwave techniques~\cite{Harty2014}; mapping both ways for two qubits would thus increase the net two-qubit gate error, but this additional mapping cost should be straightforward to reduce by improved magnetic field control. We employed the same randomized benchmarking protocol as in previous work on microwave-driven single-qubit gates~\cite{Harty2014}, here using typically 160 distinct random sequences each consisting of up to 1000 computational gates. Results are shown in figure~\ref{F:allgates}b; the average gate error is below 1\e{-3} over the entire range of gate speeds studied ($0.9\us\ldots 60\us$), with a minimum of 0.066(3)\e{-3} at a gate duration of $t_{\pi/2}=7.5\us$. This represents a five-fold reduction in error compared with previous laser-driven single-qubit gates, without incurring the overhead of composite pulse techniques~\cite{Mount2015}. For a pair of ions, we also demonstrated individual qubit addressing using the trap's axial micromotion~\cite{Navon2013a} with an estimated cross-talk error of 0.1(1)\e{-3} (see Supplemental Material). 

We now discuss the prospects for implementing the single- and two-qubit laser gates described in the present work in ion trap systems suitable for scalable quantum information processing, without sacrificing gate fidelity. Two complementary schemes are currently being pursued, the ``quantum CCD'' approach in which ions are shuttled around a large microfabricated array of interconnected traps~\cite{Wineland1998a} and a ``network'' model where multiple small traps are connected by photonic links which enable heralded entanglement~\cite{Monroe2014}. Both schemes require local deterministic gates with errors $\ll 1\%$~\cite{Nickerson2014}. Sympathetic cooling using a different ion species will also be necessary~\cite{Wineland1998a}; we have previously used \Ca{40} for this purpose~\cite{Home2009}, and have recently demonstrated that the two-qubit gate used here is also capable of mapping quantum information coherently between these two isotopes~\cite{Ballance2015}. For improved protection of logic qubits different elements can be used~\cite{Tan2015}. In the network model, a macroscopic ion trap such as that used in this work could be used at each node of the network, although it would be advantageous to have several interconnected trapping zones at each node~\cite{Nickerson2014}. In the quantum CCD model, microfabricated surface-electrode traps are likely to be necessary because of the large number of zones required~\cite{Amini2010}.

The first difficulty in using surface traps is that the ions are typically trapped much closer to the electrodes, where the electric field noise is greater, disrupting the coherent motional dynamics of the gate. For example, in the room-temperature surface trap used for the high-fidelity single-qubit work reported in~\cite{Harty2014}, we have measured motional heating and decoherence rates one to two orders of magnitude higher than in the macroscopic trap used here. According to our error model, this will limit the two-qubit gate error to a minimum of $\epsilon_g\approx 1\e{-3}$. However, significant reductions in electric field noise have been obtained in surface traps, either by cooling the electrodes to cryogenic temperatures~\cite{Labaziewicz2008} or by {\em in situ\/} cleaning of the electrode surfaces~\cite{Allcock2011,Hite2012}; this should allow significantly lower $\epsilon_g$. A second technical issue for surface traps is the proximity of the relatively powerful Raman beams to the surface: stray laser light can cause charging of the trap~\cite{Harlander2010}, which also leads to uncontrolled electric fields. We have investigated this by aligning a beam of similar power to our Raman beams on an ion trapped 75\um\ above the surface trap used in~\cite{Harty2014}; we find no detectable change in the required compensation field, at a level $\approx 1\unit{V/m}$, indicating that for light of the wavelength (397\nm) required for \Ca{43}, and these trap materials (gold electrodes on a sapphire substrate), charging is not a significant problem and there would be negligible effect on the gate error.

All the contributions to the two-qubit gate error budget (Table~\ref{T:errorbudget}) can be reduced by technical improvements; for example, improved magnetic field stabilization would reduce the spin dephasing error. The only error which is fundamentally limited is that due to photon scattering; this can be reduced to the $\approx 0.1\e{-3}$ level (in \Ca{43}) by a factor $\approx5$ increase in the Raman beam intensity \cite{Ozeri2007,Ballance2014a}. The laser power used in these experiments is modest (5\mW\ in each gate beam, with waist 27\um), and could be further reduced by integrating optical elements with the trap structure, allowing more tightly focussed beams while retaining beam-pointing stability~\cite{Mehta2015}. Solid state diode lasers are a more readily scalable technology than the frequency-doubled lasers used in this work, and it has been shown that similar power and spectral purity can be obtained from the latest generation of violet laser diodes, using optical injection locking~\cite{Schafer2015}; an array of such injected diodes could be used to implement thousands of gate operations in parallel. We conclude that, with existing technology, the 99.9\%\ two-qubit gate fidelity demonstrated here can be maintained and improved in the ion trap systems currently envisaged for the implementation of large-scale quantum computation. The task of scaling up quantum information processors beyond the present generation of few-qubit demonstrators remains a formidable one, but we hope that this work will both enable and stimulate efforts to address this technical challenge.

We note that laser-driven quantum logic gates with comparable fidelity have recently been reported by the NIST Ion Storage Group, using \Be{9} hyperfine qubits~\cite{Tan2016}. 

% approx 2500 words 

%\section*{Acknowledgements}

\vspace{1ex}
We thank other members of the Oxford ion trap group, especially D.~N. Stacey and A.~M. Steane, for their contributions to these experiments and for comments on the manuscript. We are grateful to D.~T.~C. Allcock, S.~C. Benjamin, J.~P. Home, D.~Leibfried, R.~Ozeri, T.~R. Tan and D.~J. Wineland for helpful discussions, and to the ETH Trapped Ion Quantum Information group for the development of direct digital synthesis hardware. This work was supported by the U.K.\ EPSRC ``Networked Quantum Information Technology'' Hub and the U.S.\ Army Research Office (ref.\ W911NF-14-1-0217).

% SUPPLEMENTAL MATERIAL

\section*{SUPPLEMENTAL MATERIAL} 

\subsection{Experimental apparatus and techniques}

The ions are trapped in a three-dimensional linear Paul trap with an ion-electrode distance of $500\um$. \Ca{43} ions are loaded into the trap from an enriched source (12\% $^{43}$Ca) by isotope-selective photoionization~\cite{Lucas2004} and may be held indefinitely under ultra-high vacuum conditions ($\ltish 10^{-11}\mbar$). For a single \Ca{43} ion the axial and radial trap frequencies were $f_z=1.95\MHz$ and $f_r=4.5\MHz$ respectively, the heating rate from the ground state of axial motion was measured to be $\dot{\bar{n}}=1.1 \s^{-1}$, and the coherence time for the axial motional superposition state $(\ket{n=0}+\ket{n=1})$ was found to be $\tau\gtish 200\ms$ (here and below $n$ stands for the relevant mode occupation number). Due to the finite length of the trap, there is intrinsic axial micromotion; for a two-ion crystal centred on the micromotion null the axial motion amplitude is $38\nm$, which reduces the Raman carrier Rabi frequency from $\Omega$ to $0.83\Omega$. There is a significant axial static magnetic field gradient of $\approx 0.6\unit{mG/\um}$ whose origin is not known, and which leads to a difference in the $\ket{\Downarrow}\leftrightarrow\ket{\Uparrow}$ qubit frequencies of two ions in a crystal. The global single-qubit $\pi/2$ and $\pi$ rotations used in the two-qubit gate experiments are driven by microwave signals applied to one of the trap electrodes; the Rabi frequency for the $\ket{\Downarrow}\leftrightarrow\ket{\Uparrow}$ qubit transition is $\Omega\sub{mw}/2\pi=82\kHz$. The slight difference in qubit frequencies results in imperfect microwave rotations, leading to the ``spin-echo error'' $\epsilon\sub{SE}$ (described further below).

The ions' radial and axial motion is first Doppler cooled to $\bar{n}\,\ish6$; then both the axial motional modes are cooled further by continuous Raman sideband cooling to $\bar{n}\approx 0.5$ and finally by pulsed Raman sideband cooling to $\bar{n}\approx 0.02$. Doppler cooling beams counter-propagate along the $\pi$ beam direction in figure~\ref{F:gatebeams}b. The ``gate'' qubit $(\ket{\Downarrow},\ket{\Uparrow})=(\hfslev{4S}{1/2}{4,+4}, \hfslev{4S}{1/2}{3,+3})$ and ``memory'' qubit $(\ket{\downarrow},\ket{\uparrow})=(\hfslev{4S}{1/2}{4,0}, \hfslev{4S}{1/2}{3,0})$ are initialised by optical pumping with 397\nm\ $\sigma^+$-polarized and $\pi$-polarized light respectively; $\sigma^+$ beams for preparation and readout counter-propagate along the $B$-field direction in figure~\ref{F:gatebeams}b. Both types of qubit are read out by state-selective shelving to the \lev{3D}{5/2} level followed by manifold-selective fluorescence detection~\cite{Myerson2008}. We detect the total fluorescence using a photomultiplier; for two ions, this does not allow us to distinguish the $\ket{\Downarrow\Uparrow}$ state from $\ket{\Uparrow\Downarrow}$, but this is not a limitation in these experiments (see below).

The sideband cooling, single-qubit gate and two-qubit gate operations are driven by three different pairs of Raman laser beams, as illustrated in figure~\ref{F:gatebeams}. The beams for the sideband cooling and the two-qubit gate propagate at $45\degree$ and $135\degree$ to the trap $z$-axis, such that their difference $\bm{k}$-vector lies along $z$, giving a Lamb-Dicke parameter for the two-ion centre-of-mass mode $\eta=0.123$. The beams for the single-qubit gates copropagate ($<10\mrad$ beam angle) giving negligible coupling to the motion ($\eta<10^{-3}$). The $\approx397\nm$ Raman beams are derived from a pair of frequency-doubled diode lasers, the second (slave) of which is injection-locked at 794\nm\ from the first (master) via a double-pass $800\MHz$ acousto-optic modulator (AOM), allowing the 3.2\GHz\ qubit splitting to be bridged for qubit carrier and sideband manipulations~\cite{Linke2013a}. The output beam from the master is split, switched using two AOMs, and brought close to the trap using $\approx 1\m$ long optical fibres; these are the beams used to drive the two-qubit gate. A third Raman beam is derived from the slave in a similar way, and is used in combination with a beam from the master laser to drive either single-qubit gates or sideband cooling (the two alternative slave beam paths are both shown in figure~\ref{F:gatebeams}b).

\begin{figure}[t]
\centering
\includegraphics[width=0.9\onecolfig]{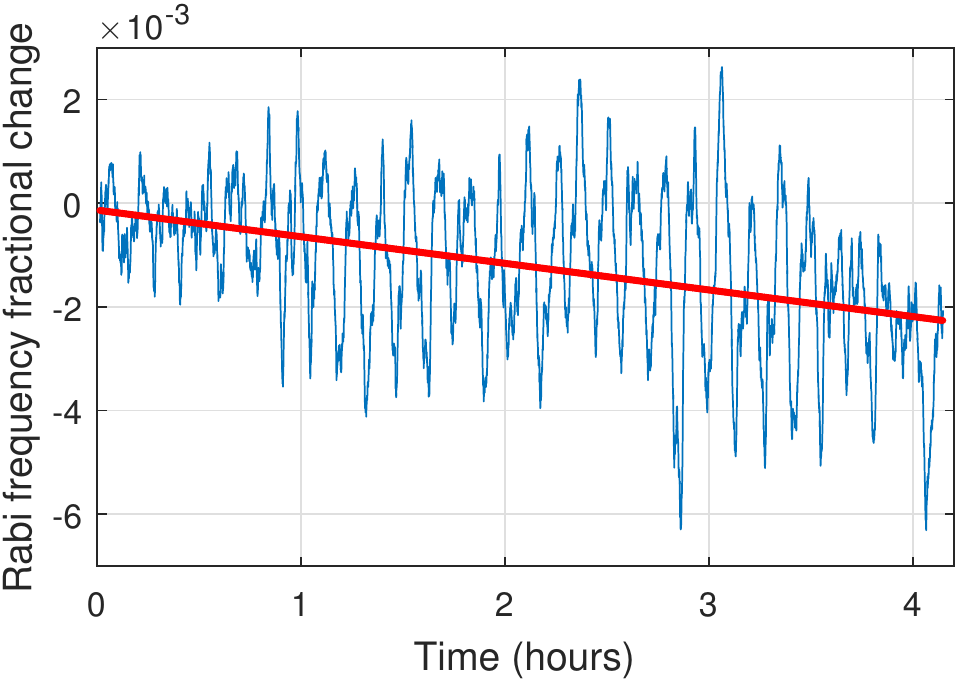}
% CJB figure 6.11
\caption{Fractional drift in Rabi frequency for single-ion qubit carrier transitions driven by Raman beams aligned along the two-qubit gate beam paths. At time zero the beam pointing was optimized to give maximum Rabi frequency. The periodic fluctuations of $\ish\pm2\e{-3}$ are synchronous with the ($\approx$\,8 minute) laboratory air-conditioning cycle. A straight-line fit (red line, slope $-0.5\e{-3}$/hour) is shown to give an indication of slow drift. A fractional change of $\pm5\e{-3}$ would give a two-qubit gate error $0.06\e{-3}$, hence these fluctuations are sufficiently small that they do not contribute significantly to the gate error (Table~\ref{T:errorbudget}).}
\label{F:beamdrift}
\end{figure}

The Raman beams are focused to waists of $w=27\um$ ($1/e^2$ intensity radius) at the ions. The power used in each beam was up to $5\mW$. Power drifts after the fibres are stabilized by feedback to each AOM's r.f.\ drive power once every 20\ms\ experimental sequence, before the qubit state preparation. The gate beams are steered onto the ions using home-made piezo-actuated mirror mounts, by maximizing the qubit carrier Rabi frequency $\Omega$ for a single ion; this aligns the beams on the trap centre with a precision of $\pm 0.5\um$. We characterize long-term intensity and beam pointing noise by measuring the drift of $\Omega$, and find $\delta\Omega/\Omega<0.5\%$ over periods of several hours (figure~\ref{F:beamdrift}); this test was performed with non-copropagating beams, using the two-qubit gate beam paths. This drift would lead to systematic single- and two-qubit gate errors of $<0.06\e{-3}$. Fast intensity noise was measured directly at the fibre outputs (typically 0.2\% r.m.s.), and relative phase noise was investigated by optical heterodyne measurements between both master/master and master/slave Raman beam pairs~\citesec{Ballance2014a}{6.4.2}. Amplitude and phase noise are estimated to contribute negligibly to the two-qubit gate error, but we believe the latter is the dominant source of error for the single-qubit gates~\cite{Harty2013,Ballance2014a}. The polarization of the $\sigma^{\mp}$ Raman beam is optimized with a $\lambda/4$ waveplate to null any differential single-qubit light shift arising from this beam. 

The Raman AOMs are driven by direct digital synthesis (DDS) radiofrequency sources, which allow dynamic control of the beam intensities and frequencies during the experimental sequence. The frequency control is used to switch the sideband cooling between the centre-of-mass and stretch mode frequencies. We use the amplitude control to shape the intensity turn-on and turn-off of one of the Raman gate beams with a characteristic time of 1.5\us\ (and ensure the other beam is turned on before, and off after, the pulse-shaped beam). We estimate that without this pulse-shaping, there would be an additional average two-qubit gate error of up to $4\e{-3}$ at $t_g=100\us$ due to off-resonant carrier light shifts (see below).

The ion spacing is set by applying the Raman gate beams to drive resonantly either the centre-of-mass mode ($\delta=f_z$) or the stretch mode ($\delta=f_z\sqrt{3}$), for an initial state $\ket{\Downarrow\Downarrow}$ with both motional modes cooled to the ground state. When the ion spacing is a half-integer number of standing wave periods, there is maximal excitation of the stretch mode and minimal excitation of the centre-of-mass mode, because the force on the two ions is in opposing senses. The motional excitation is probed using a Raman red-sideband $\pi$ pulse. 

We optimize the gate duration $t_g$ by scanning the length of the Raman gate pulses, as shown in figure~\ref{F:parity}b, and minimizing the `single spin-flip' signal $(P_{\Downarrow\Uparrow}+P_{\Uparrow\Downarrow})$, as measured at the end of the spin-echo sequence. Setting $t_g$ to a precision of 0.1\% should lead to a systematic gate error of $\epsilon_g\approx 0.1\e{-3}$. Once the gate duration has been set, we optimize the accumulated geometric phase (i.e.\ the total area swept out in the motional phase space) by adjusting the power in one of the Raman beams to equalize the populations $P_{\Downarrow\Downarrow}$ and $P_{\Uparrow\Uparrow}$, again measured at the end of the spin-echo sequence. We aim to balance the populations to a precision of 1\%, which gives a systematic gate error of $\epsilon_g\approx 0.1\e{-3}$.

Further details of the experimental apparatus, characterization experiments and methods are given in~\cite{Linke2012a,Ballance2014a}.

\subsection{Measuring the two-qubit gate error}

We measure the two-qubit gate error by using the two-qubit gate, along with single-qubit rotations, to generate the Bell state $\ket{\psi_+}=(\ket{\Downarrow\Downarrow}+\ket{\Uparrow\Uparrow})/\sqrt{2}$, whose fidelity we measure with a partial tomography procedure~\cite{Sackett2000}. The Bell state error includes errors from the qubit state-preparation and measurement (SPAM), the single-qubit rotations, and the two-qubit gate itself. In our reported gate errors we have corrected for the SPAM error $\epsilon\sub{SPAM}$ and the dominant single-qubit error $\epsilon\sub{SE}$, neither of which is intrinsic to the gate operation, in order to obtain the best estimate of the error in the two-qubit gate itself. This allows quantitative comparison both with our own error model and with experimental gate errors measured by other techniques which naturally exclude SPAM and single-qubit errors, such as randomized benchmarking~\cite{Gaebler2012a,Barends2014}. We describe in this section our methods to determine $\epsilon\sub{SPAM}$ and $\epsilon\sub{SE}$, as well as our data fitting procedure.

For a single \Ca{43} ion, we measure a combined state preparation and single-shot readout error of 0.90(5)\e{-3} using state-selective shelving and thresholded fluorescence detection~\cite{Myerson2008}. For two ions, the SPAM error per qubit $\epsilon\sub{SPAM}$ is greater as we use a longer detection period (1.9\ms), to improve discrimination between 0, 1 and 2 ions fluorescing, which increases the error due to spontaneous decay from the \lev{3D}{5/2} shelf (lifetime 1168(7)\ms~\cite{Barton2000}). The two photon-count thresholds were set before the gate runs using independent control data, and were not adjusted in data analysis. We measure $\epsilon\sub{SPAM}$ with a two-ion crystal by preparing and reading out the $\ket{\Downarrow\Downarrow}$ and $\ket{\Uparrow\Uparrow}$ states typically $8\e{4}$ times each, allowing us to extract the readout errors for the $\ket{\Downarrow}$, $\ket{\Uparrow}$ qubit states and hence the average per-qubit error $\epsilon\sub{SPAM}=\frac{1}{2}(\epsilon_\Downarrow+\epsilon_\Uparrow)$. We then construct a linear map for two qubits to infer the populations of the states $\ket{\Downarrow\Downarrow}, (\ket{\Downarrow\Uparrow}$ or $\ket{\Uparrow\Downarrow}), \ket{\Uparrow\Uparrow}$ from the experimental signals (0, 1 or 2 ions fluorescing)~\citesec{Ballance2014a}{C.2}. On the day when the $t_g=100\us$ two-qubit gate data were taken, we measured $\epsilon\sub{SPAM}=1.74(7)\e{-3}$ using two ions. To investigate the stability of the SPAM error, we made two-ion SPAM measurements on six separate days with similar precision; these measurements have a standard deviation of 0.11\e{-3}, somewhat in excess of the statistical precision, but small compared with the uncertainty in $\epsilon_g$. If we did not correct for SPAM errors, the apparent infidelity in the Bell state would increase by $\approx 3\epsilon\sub{SPAM}$ (see~\citesec{Ballance2014a}{C.4}).

In calculating the SPAM error for the states $\ket{\Downarrow\Uparrow}$ and $\ket{\Uparrow\Downarrow}$ we make the assumption that it is identical for both ions. We confirm that the state-preparation and state-selective shelving error is identical for both ions, by measuring the SPAM error for a single ion placed at each ion location of the two-ion crystal. However, for two-ion experiments we expect there to be a small difference in the fluorescence detection error between the states $(\ket{\Downarrow\Uparrow},\ket{\Uparrow\Downarrow})$ and the states $(\ket{\Downarrow\Downarrow},\ket{\Uparrow\Uparrow})$ due to spontaneous decay from \lev{3D}{5/2} during the fluorescence detection period. A numerical model of the readout process including shelf decay shows that, for our typical bright and dark count rates, our method of estimating the SPAM error gives a systematic overestimate of gate errors of $\approx 0.1\e{-3}$; this is substantially smaller than the statistical uncertainty in the gate error and we do not attempt to correct for it.

As a result of the axial magnetic field gradient, for two ions the $\ket{\Downarrow}\leftrightarrow\ket{\Uparrow}$ qubit frequencies differ by $\delta f=4.91\kHz$. Although this does not affect the gate operation itself, it does introduce errors in the global single-qubit $\pi/2$ and $\pi$ rotations in the spin-echo sequence. The Rabi frequency for the single-qubit operations is $\Omega\sub{mw}/2\pi=82\kHz$, thus the errors due to the $\pm \frac{1}{2}\delta f$ detuning errors for single $\pi$ pulses and $\pi/2$ pulses are $\approx 1\e{-3}$ and $\approx 0.1\e{-3}$ respectively. In the spin-echo sequence these errors coherently add or subtract, depending on the phase of the pulses relative to the precessing qubit phases; as we increase the length of the spin-echo sequence to accommodate slower gates the relative phases of the pulses change due to the different qubit detunings. To quantify the effect of this error, we perform numerical simulations to determine the error in producing a Bell state from $\ket{\Downarrow\Downarrow}$ assuming a perfect two-qubit phase gate~\citesec{Ballance2014a}{8.3.2}. We find the error $\epsilon\sub{SE}$ varies sinusoidally as a function of the duration of the spin-echo sequence, with a first maximum of $1.8\e{-3}$ when $t_g\approx200\us\approx 1/\delta f$. As this error depends only on the qubit frequency difference $\delta f$ and the Rabi frequency $\Omega\sub{mw}$, both of which we can determine accurately, negligible uncertainty in $\epsilon_g$ is introduced by subtracting $\epsilon\sub{SE}$ from the Bell state error. There are several other sources of error from the single-qubit operations, such as detuning error of the $\frac{\pi}{2}[\phi]$ tomography pulses, imperfectly set pulse areas and off-resonant excitation; these are estimated to be $\ltish 0.1\e{-3}$, well below the statistical error in $\epsilon_g$, and we do not attempt to correct for them.

We use maximum-likelihood fitting to fit the parity fringes (figure~\ref{F:parity}a) with the function $C_0 + C\sin{[2(\phi-\phi_0)]}$ where the phase offset $\phi_0$ allows for an undetermined phase shift due to drift in the applied magnetic field $B$. We find that $\phi_0$ differs by 0.8(2)\degree\ for the two $t_g=100\us$ data sets shown in figure~\ref{F:parity}a, which may be accounted for by a drift $\delta B\approx 0.01\uT$ during the \mbox{$\approx 30\min$} period between the two runs, consistent with the level we typically observe. Such drift could be straightforwardly eliminated by feedback to the magnetic field, for example using a field sensor or co-trapped sympathetic cooling ions of a second species. The fitted amplitude offset $C_0=0.000(1)$ is consistent with zero. A joint fit to both data sets gives a fitted amplitude $C=0.9953(11)$; together with the population measurement without tomography pulses, $P_{\Downarrow\Downarrow}+P_{\Uparrow\Uparrow}=0.9997(8)$, we obtain the Bell state fidelity $F=\frac{1}{2}(C+P_{\Downarrow\Downarrow}+P_{\Uparrow\Uparrow})=0.9975(7)$, before correction for $\epsilon\sub{SE}$.

The statistical errors on the parity signal are distributed according to a binomial distribution, as the parity is either even or odd. Hence it is appropriate to fit the data using a maximum-likelihood method assuming binomially-distributed measurements (as opposed to, for example, least-squares fitting, which assumes Normally-distributed data~\citesec{NumRec}{15}). To verify our fitting procedure we generate many synthetic sets of parity fringes, using binomially-distributed random numbers and the same number of trials as in the experiments, for a range of contrasts $0.985<C<0.999$. We find that the maximum-likelihood fit gives an unbiased estimate of the contrast used to generate the synthetic data sets. We also find that using a least-squares fit would lead to a systematic under-estimate of the Bell state error (by $\approx 1\e{-3}$ for our typical experimental parameters). Further details of these tests are given in~\citesec{Ballance2014a}{8.3.4}.

\subsection{Sources of error in two-qubit gate experiments}

We consider the dominant sources of error in the two-qubit gate, as listed in table~\ref{T:errorbudget} and plotted in figure~\ref{F:allgates}a. A more detailed discussion may be found in~\citesec{Ballance2014a}{4.4}, where the following sources of error are also considered: unequal illumination of the two ions by the gate beams, incorrect spacing of the ions, amplitude and phase noise in the Raman beams, drifts in the axial mode frequency, and the effect of coupling between radial and axial motional modes. For our conditions, these errors are estimated to be negligible compared with those discussed here.

\begin{figure}[t!]
\flushleft{(a)}
\center\includegraphics[width=0.9\onecolfig]{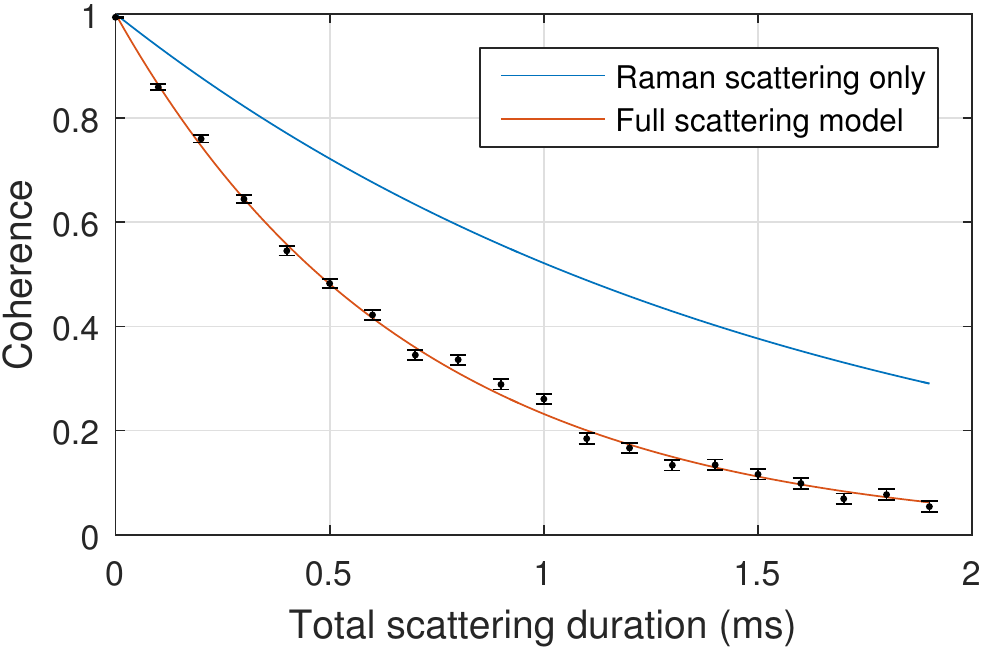}
\flushleft{(b)}
\center\includegraphics[width=0.9\onecolfig]{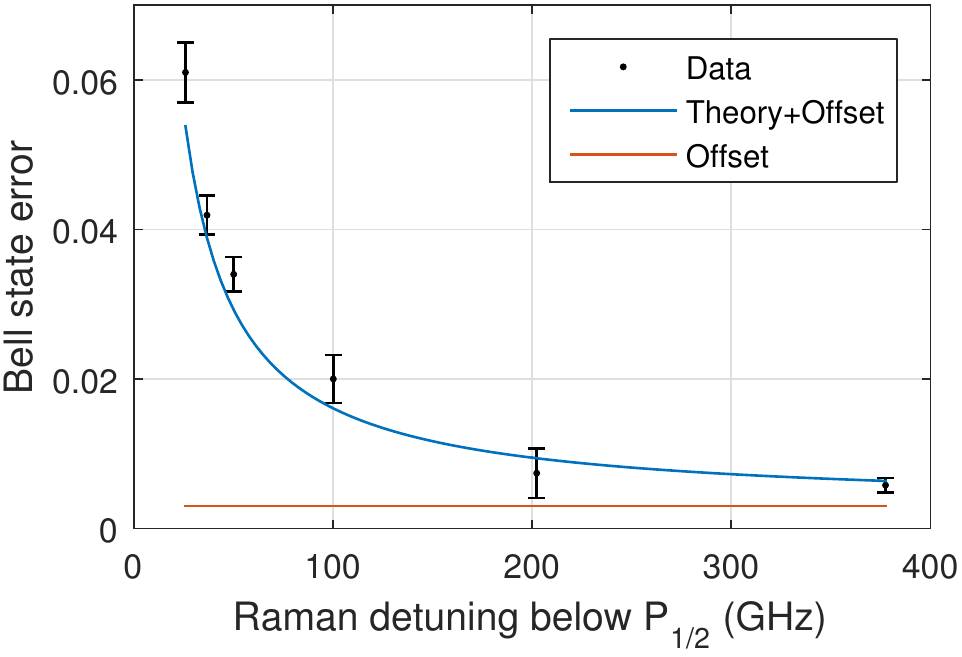}
% CJB figure 3.6
\caption{Measurements of photon scattering errors. 
(a) Decoherence of a $(\ket{\Downarrow}+\ket{\Uparrow})/\sqrt{2}$ superposition state of a single ion due to photon scattering from a single $\sigma^\pm$ polarized Raman beam. This was measured by applying scattering pulses during the gaps in a CPMG Ramsey sequence (used to suppress decoherence due to magnetic field fluctuations~\cite{Ozeri2005}), and observing the decay in Ramsey fringe contrast. The blue line is the model prediction if the only source of decoherence is Raman photon scattering. The red line is the model including dephasing from Rayleigh scattering~\cite{Uys2010}, and agrees well with the measured data. The Raman beam power was 5\mW\ and the detuning was $\Delta=-138\GHz$. 
(b) Measurement of Bell state error from photon scattering. The gate duration is fixed at $t_g=39\us$, the Raman detuning $\Delta$ is varied, and the Raman beam power adjusted so as to keep the Rabi frequency constant. The only change in the gate error is from the change in the photon scattering error. The blue line is the theoretical scattering model with an offset (2.6\e{-3}, red line) added to account for the remaining error sources. The model includes the effect of the intrinsic axial micromotion, which increases the laser intensity required to achieve a given Rabi frequency, thus increasing the scattering error. 
}
\label{F:scattering}
\end{figure}

As shown in figure~\ref{F:allgates}a, the dominant contribution to the error for gate times $3\us \ltish t_g \ltish 100\us$ is due to photon scattering. This comprises both Raman and Rayleigh contributions; the latter also contributes to decoherence because of the differential scattering rates from the two qubit states~\cite{Uys2010}. We adapted the scattering theory of previous authors~\cite{Ozeri2007,Uys2010} for our situation and checked the predictions by: (i) measuring the coherence decay of a single-qubit $(\ket{\Downarrow}+\ket{\Uparrow})/\sqrt{2}$ superposition state while a single Raman beam is applied for a variable duration (figure~\ref{F:scattering}a); (ii) applying the two-qubit gate and measuring the Bell state error as a function of $\Delta$, while adjusting the Raman beam powers so as to keep the Rabi frequency constant (figure~\ref{F:scattering}b). We calibrated the beam intensities by auxiliary experiments with single ions, thus ensuring these tests are parameter-free. Both experiments agree well with the theoretical scattering calculations, within the statistical uncertainty of the data. 

It would be advantageous to work at a static magnetic field of $B=14.6\mT$, which gives access to a \Ca{43} magnetic field-independent hyperfine memory qubit~\cite{Harty2014}. The two-qubit phase gate can still be applied to the ($\ket{\Downarrow},\ket{\Uparrow}$) states without modification at this field, but the photon scattering error is altered because of the slightly different composition of the atomic states due to hyperfine state-mixing. We calculate that this would increase the total photon scattering error by $\approx 5\%$.

% spin dephasing

For gate times $t_g \gtish 200\us$ the dominant error is spin dephasing from magnetic field fluctuations. Although the spin-echo sequence protects against slow fluctuations in magnetic field, noise that is uncorrelated between the two halves of the spin-echo will lead to spin dephasing. To characterize this, we measure for a single ion the fringe contrast of a spin-echo sequence without gate pulses, scanning the phase of the final $\pi/2$ pulse, as a function of the length $t\sub{SE}$ of the spin-echo. The contrast decay is well represented by an empirical model quadratic in $t\sub{SE}$. As the two-qubit gate operation commutes with the spin-dephasing operator, this contrast decay may be modelled as an effective qubit readout error~\citesec{Ballance2014a}{8.3.3}; its contribution to the gate error is plotted on figure~\ref{F:allgates}a.

% light shift error

The travelling standing wave resulting from the interference of the Raman gate beams necessarily gives rise to a differential light shift (a.c.\ Stark shift) on each qubit with an amplitude that oscillates at the Raman difference frequency $\delta$. Over the course of the gate operation this light shift adds phase shifts to the qubits that depend on the (uncontrolled) optical phase difference of the Raman beams. If these phase shifts are an exact multiple of $2\pi$, there is no error (giving periodic minima for this error as a function of $t_g$), but otherwise they reduce the fidelity of the gate operation. We greatly reduce this source of error by shaping the laser pulses as described above. Off-resonant excitation of the axial stretch mode of motion is negligible $(\ish \eta^2)$ compared with this `carrier light shift' effect, and it is likewise suppressed by the pulse-shaping. The red curve on figure~\ref{F:allgates}a shows this light shift error, taking into account the nominal pulse shape.

% motional temperature, heating and dephasing 

We finally consider three sources of error associated with the ions' motional states, due to: (i) non-zero temperature of the ions; (ii) ambient heating of the motion during the gate; and (iii) dephasing of the motional mode during the gate. 

\begin{figure}[t!]
\flushleft{(a)}
\center\includegraphics[width=0.9\onecolfig]{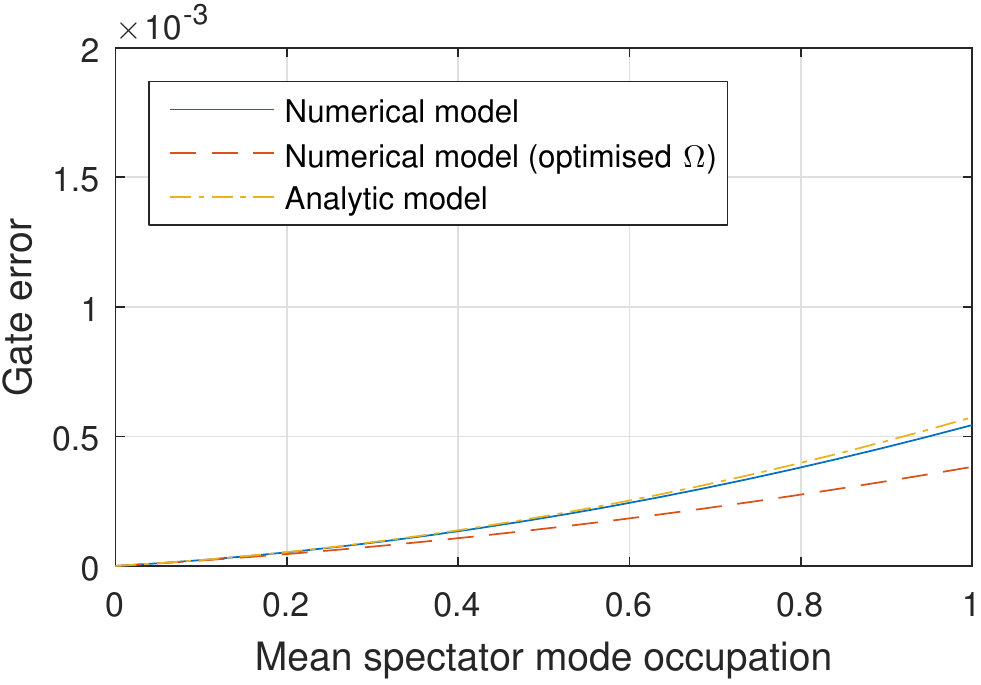}
\flushleft{(b)}
\center\includegraphics[width=0.9\onecolfig]{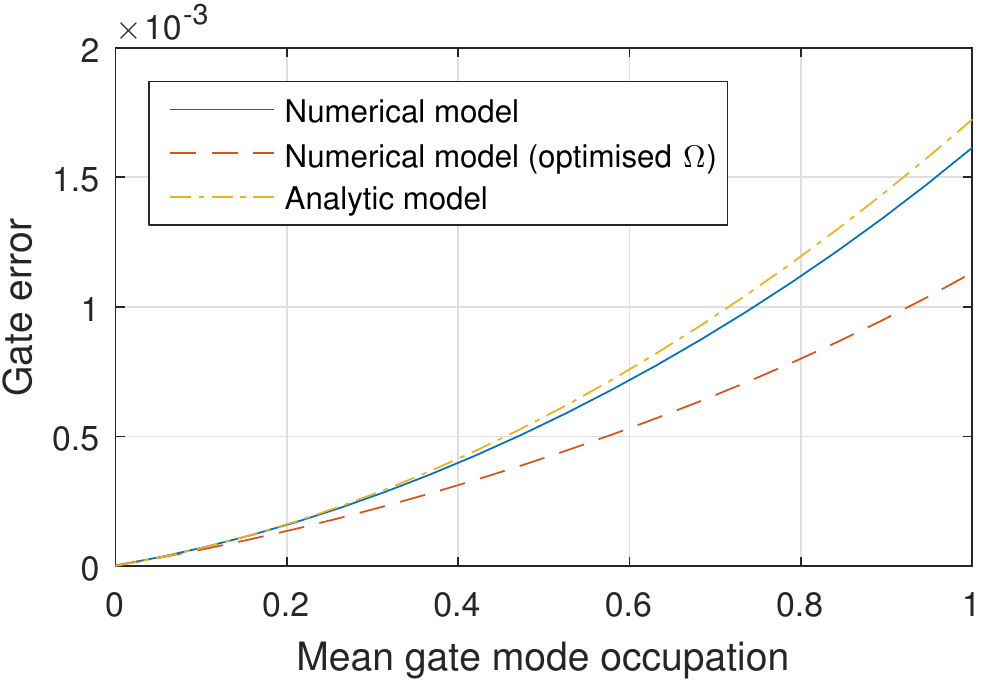}
\caption{Analytic and numerical models of the gate error due to finite temperature of the axial ion motion. 
% CJB figure 4.1
(a) Gate error as a function of mean thermal occupation $\bar{n}$ of the spectator mode with $\eta=0.094$, corresponding to the two-ion stretch mode for a \Ca{43} crystal with centre-of-mass frequency $f_z=1.93\MHz$. The analytic and numerical model are for a gate with Rabi frequency set assuming $\bar{n}=0$; the `optimised $\Omega$' numerical result is the minimum error achievable after empirically optimising the gate Rabi frequency at each value of $\bar{n}$. 
% CJB figure 4.2
(b) Gate error as a function of mean thermal occupation $\bar{n}$ of the gate mode with $\eta=0.12$, corresponding to the two-ion centre-of-mass mode for a \Ca{43} crystal with $f_z=1.93\MHz$. The numerical model is for a gate with Rabi frequency set assuming $\bar{n}=0$; the analytic model is the result derived for the {\em spectator\/} mode, which also appears to describe this error well. The `optimised $\Omega$' numerical result is the minimum error achievable after empirically optimising the gate Rabi frequency at each value of $\bar{n}$.
}
\label{F:temperror}
\end{figure}

(i) The two-qubit gate we implement is insensitive to the ions' motion to first order in the Lamb-Dicke parameter~\cite{Leibfried2003} but for $\eta\,\ish 0.1$ we need to consider higher-order terms. Thermal occupation of both the centre-of-mass ``gate'' mode ($\eta=0.12$), and the stretch ``spectator'' mode ($\eta=0.094$) introduce errors. Motional excitation of the spectator mode simply reduces the coupling strength of the ions to the laser field; averaging over a thermal state with mean occupation number $\bar{n}$, and neglecting heating during the gate, we find the contribution to the gate error is given by~\citesec{Ballance2014a}{4.4.4}
\[
\epsilon_{\bar{n}} = \wee{1}{4} \pi^2 \eta^4 \bar{n} (2\bar{n} + 1)
\]
We compare this approximate analytic expression with a numerical result in figure~\ref{F:temperror}a, and find that both models agree well for $\bar{n}<1$. We also numerically model the error due to excitation of the gate mode itself; this situation is more complex as the motional mode evolves during the gate, but we observe that the above expression, with the appropriate $\eta$, nevertheless models the resulting gate error well for $\bar{n}<1$, figure~\ref{F:temperror}b. We conclude that the error due to the finite gate mode temperature is significantly larger, and that for our typical mode temperatures of $\bar{n}<0.05$, the total contribution to $\epsilon_g$ is $<0.04\e{-3}$.

(ii) Heating of the gate motional mode during the gate operation leads to error. Following ref.~\cite{Turchette2000}, we model the heating as a coupling to an infinite temperature bath. From a numerical model, we find that the gate error $\epsilon_h$ due to heating is, for $\epsilon_h\ll 0.1$, given by
\[
\epsilon_h = \frac{\dot{\bar{n}} t_g}{2K}
\]
where $\dot{\bar{n}}$ is the heating rate of the gate mode and $K$ the number of loops in phase space described by the motion during the gate ($K=2$ for our gate with $t_g=2/\delta_g$). This error decreases as $K$ increases because the motional excursion in phase space is smaller; in the limit of large $K$ the motional excitation becomes completely virtual and the mechanism is insensitive to motion, as in the original scheme for the related $\sigma_x\otimes\sigma_x$ gate~\cite{Sorensen1999}. Taking the heating rate of the two-ion centre-of-mass gate mode to be $\dot{\bar{n}}=2.2\persec$ (i.e.\ twice that measured for a single ion), we obtain $\epsilon_h=0.06\e{-3}$ at $t_g=100\us$.

(iii) Dephasing of the motional gate mode while the spin state is entangled with the motional state, as it is during the gate operation, also leads to error. We model the dephasing of the gate motional mode with a Lindblad operator $L=a^\dagger a \sqrt{2/\tau} $ where $a$ is the gate mode annihilation operator and $\tau$ is the motional decoherence time~\cite{Turchette2000}. $L$ causes the coherence of a motional superposition state $\ket{n}+\ket{n'}$ to decay at a rate $(n-n')^2 / \tau$. Integrating the master equation we find that the gate error due to motional dephasing is described by
\[
\epsilon_d = \alpha_K \frac{t_g}{\tau}
\]
where, for $K=\{1,2,4\}$, the numerical coefficient $\alpha_K=\{0.686,0.297,0.137\}$. As expected, $\epsilon_d$ decreases with $K$, similarly to the heating rate error. For $K=2$ and $\tau=200\ms$ we find $\epsilon_d=0.15\e{-3}$ at $t_g=100\us$. This dephasing error is thus the dominant contribution to the motional error plotted in figure~\ref{F:allgates}a.

\subsection{Single-qubit gate experiments}

For the single-qubit addressing demonstration we positioned one ion of a two-ion crystal on the axial micromotion null. The Raman Rabi frequency on the first micromotion sideband of the $\ket{\Downarrow}\leftrightarrow\ket{\Uparrow}$ qubit transition (driven by the $\pi$ and $\sigma^{\mp}$ beams, with $\delta=f_0-7f_B-f\sub{mm}$ where the micromotion frequency $f\sub{mm}=30.0\MHz$) was measured to be $\Omega\sub{mm}/2\pi=36.5(2)\kHz$ for the off-null ion and $\Omega'\sub{mm}/2\pi=0.2(1)\kHz$ for the on-null ion. With this Rabi frequency ratio a $\pi$-pulse resonant with the micromotion sideband on the off-null ion would induce an error of $(\pi^2/4)(\Omega'\sub{mm}/\Omega\sub{mm})^2 \approx 0.1(1)\e{-3}$ on the on-null ion~\cite{Warring2013a}. Other off-resonant processes that would drive transitions for the on-null ion are estimated to be negligible.

For the single-qubit randomized benchmarking experiment, we used the ``atomic clock'' memory qubit $(\ket{\downarrow},\ket{\uparrow})=(\hfslev{4S}{1/2}{4,0}, \hfslev{4S}{1/2}{3,0})$, whose coherence time was measured to be $T_2^*=6(1)\s$ at $B=0.196\mT$. This qubit is prepared by optical pumping with 397\nm\ $\pi$-polarized light, and read out in the same way as the $(\ket{\Downarrow},\ket{\Uparrow})$ qubit. SPAM errors are significantly larger ($\approx 3\%$) for this simple preparation and readout method~\cite{Szwer2009}, but this is not a limitation here. For these experiments, the ``slave'' beam (figure~\ref{F:gatebeams}b) was aligned along the $\sigma^\pm_\parallel$ beam path instead of the $\pi$ beam path, as sideband cooling is not necessary.

\begin{figure}[t]
\center\includegraphics[width=0.9\onecolfig]{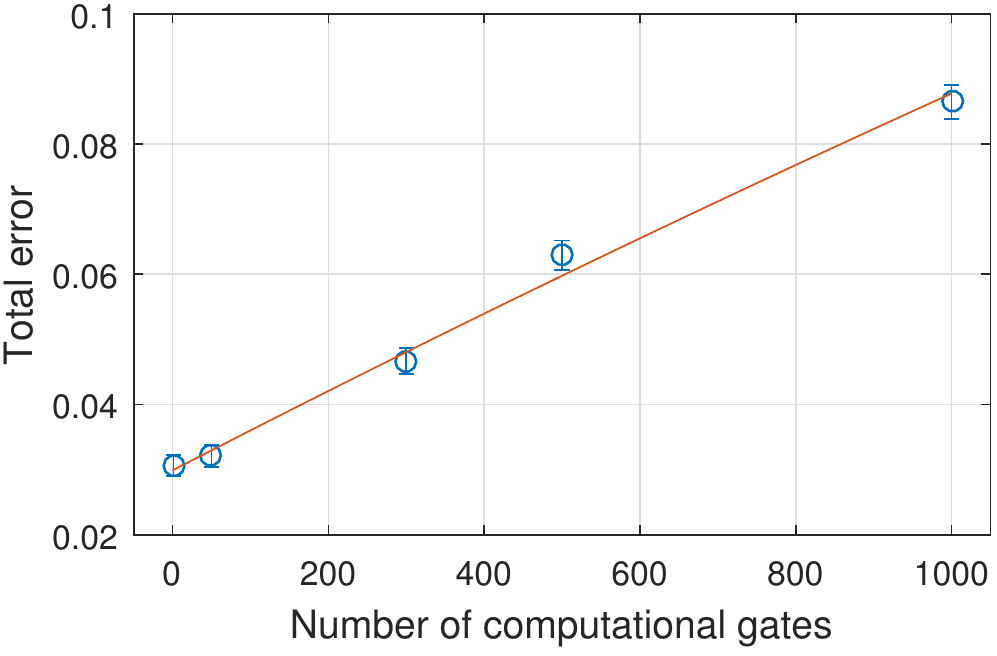}
% CJB figure 7.7
\caption{Single-qubit randomized benchmarking experiment. The plot shows the randomized benchmarking data set for the optimum single-qubit gate time $t_{\pi/2}=7.45\us$. The fitted curve gives an error per computational gate of $0.066(3)\e{-3}$. Each computational gate consists of, on average, two physical $\pi/2$ pulses. For each of the 5 sequence lengths shown, 32 distinct random sequences were used, each of which was repeated 300 times. The error bars are $1\sigma$ statistical errors. 
}
\label{F:RBMexpt}
\end{figure}

For laser-driven single-qubit gates on the memory qubit the coupling to the neighbouring hyperfine $M_F$ states is strongly suppressed for the nominal $\sigma^\pm$ and $\sigma^\mp$ Raman beam polarizations, leading to negligible off-resonant excitation ($\ish 10^{-8}$ gate error). For a $\pi$ polarization impurity of 1\%, corresponding to a 10\mrad\ alignment error between the beam direction and the quantization axis, the gate error due to off-resonant excitation would be $\ish 0.01\e{-3}$ at $t_{\pi/2}=0.5\us$; this error decreases quadratically with $t_{\pi/2}$. Hence in these experiments it is not necessary to shape the laser pulses. Diagnostic experiments show that the Rabi frequency amplitude noise, gate detuning error, differential qubit light shift, and pulse area error each contribute an error-per-gate of $\ltish 0.01\e{-3}$. The dominant source of error for all gate times is believed to be differential phase noise between the two Raman beams; Monte-Carlo simulations of the benchmarking experiments using the measured phase noise show qualitative agreement with the experimental data for both fast and slow gates. We note that the single-qubit gates are driven by a pair of beams with $3.226\GHz$ splitting, with one beam taken from the master laser, and one from the slave: these beams are measured to have much larger relative phase noise than the two beams with $\approx 2\MHz$ difference frequency used to drive the two-qubit gate, both of which are sourced from the master laser. An example randomized benchmarking data set is shown in figure~\ref{F:RBMexpt}. Further discussion of the error sources is given in~\citesec{Ballance2014a}{7.3}.

% Supplementary Information: approx 3500 words

%\section{References}

%\bibliography{libraryClean}

% Manually edited BBL: several refs addded by hand (DML)

\end{document}